\documentclass[a4paper,10pt,oneside]{article}
\usepackage{icad2018a,amsmath,epsfig,times,url,gensymb, csquotes,hyperref, authblk}
\usepackage{multirow}
\usepackage{subfig}

\begin{document}

\title{\large \bf Sonifying stochastic walks on biomolecular energy landscapes}
\author[1,*]{Robert E. Arbon}
\author[1,2,3,*]{Alex J. Jones}
\author[1,4]{Lars A. Bratholm}
\author[3]{Tom Mitchell}
\author[1,2]{David R. Glowacki}
\affil[*]{{\small These authors contributed equally to this manuscript}}
\affil[1]{{\small School of Chemistry \\ University of Bristol \\ Bristol, BS8 1TS, UK }}
\affil[2]{{\small Dept. of Computer Science \\ University of Bristol \\ Bristol, BS8 1UB, UK}}
\affil[3]{{\small Dept. of Computer Science and Creative Technologies\\University of the West of
England\\
Bristol, BS16 1QY, UK}}
\affil[4]{{\small School of Mathematics, University of Bristol, 
University Walk,
Bristol,
BS8 1TW, UK}}
\affil[ ]{\small\texttt {\{robert.arbon, alex.j.jones, lars.bratholm, glowacki\}\small @bristol.ac.uk}}
\affil[ ]{\texttt{tom.mitchell@uwe.ac.uk}}
\date{}

\ninept
\maketitle

\begin{sloppy}

\begin{abstract}
Translating the complex, multi-dimensional data from simulations of biomolecules to intuitive  knowledge is a major challenge in computational chemistry and biology. The so-called ``free energy landscape" is amongst the most fundamental concepts used by scientists to understand both static and dynamic properties of biomolecular systems. In this paper we use Markov models to design a strategy for mapping features of this landscape to sonic parameters, for use in conjunction with visual display techniques such as structural animations and free energy diagrams.  
\end{abstract}

\section{Introduction}

\label{sec:intro}
Richard Feynman famously stated \cite{Feynman} that ``everything that living things do can be understood in terms of the jigglings and wigglings of atoms''.  A complete understanding of how these atomic jigglings and wigglings give rise to the structure, dynamics and  function of biomolecules remains an outstanding scientific challenge with implications across a wide range of disciplines. For example, dynamical processes like protein folding are implicated in neurological diseases (e.g. Alzheimer's) \cite{Hashimoto2003} and conformational changes in enzymes are linked to their biological function \cite{Singh2015}.  

Computer simulation is an important tool to understand biomolecular dynamics because of its ability to reveal chemical information at the atomic level  with a high degree of temporal and spatial resolution \cite{Adcock2006,Antoniou2006}. Its popularity is associated with three developments: accurate and computationally efficient ways of modeling the interactions between atoms (the atomic `force-field')\cite{Kamp2008, Mackerell2004}, the increasing availability of highly parallel computer architectures such as general purpose graphical processing units (GP-GPUs)\cite{Stone2010}, and a variety of user friendly software packages which exploit both these developments \cite{Openmm4, amber}. This has enabled the study of bigger systems at longer timescales, moving the dynamics of biomolecular systems into the `big-data' era \cite{Graaf2015}. 

Extracting scientific information from the output of computer simulations is difficult owing to the quantity of data available. Output from molecular dynamics (MD) simulations \cite{alder1959} include time series of atomic positions (known as \emph{trajectories}) and associated data (e.g., system energy, volume, pressure, etc.). Making sense of this data requires reducing the data dimensionality by removing irrelevant features and producing an accurate but understandable model of the process being investigated.  

Analyzing trajectories is commonly performed through visual display using animations of the molecules (often with atoms rendered as balls and chemical bonds as sticks). However, there is typically far too much data for a researcher to understand. Dimensionality reduction is achieved by only displaying certain atoms while features of the data can be calculated and mapped to visual aesthetics. For example, common structural motifs in proteins, such as alpha-helices, can be drawn as a cartoon helix on top of the molecular structure to highlight their presence.  Another example is to map the colour of the rendering to the degree of conformational flexibility of a particular part of the molecule. 

A particularly important feature of the system is its \emph{free energy}. Any given molecular configuration has an associated free energy, from which several important properties can be calculated \cite{Atkins}. Often described as a ``landscape'', it is commonly represented as a topographical contour map. ``Mapping'' the free energy landscape of biomolecular systems remains a significant challenge. Nevertheless, understanding how a molecule's 3D structure relates to its free energy landscape is crucial for scientists to gain an understanding of biomolecular dynamics. The software package MolPX \cite{molpx} has attempted this by linking two separate visual objects - molecular animations and free energy diagrams. However this strategy has two drawbacks: (1) the free energy landscape is limited to two dimensions and (2) the researcher's focus is split across the two visual objects. Display of higher dimensional landscapes is possible by combinations of two dimensional projections but this only exacerbates problem (2).   

Sonification has the ability to overcome problems with displaying high dimensional free energy landscapes: the topography and important features of the landscape can be heard concurrently with visual structural information. However, creating a sonic representation of features of the free energy landscape presents a number of technical and design related challenges which are explored in this paper.  

Two major techniques for approaching an auditory display challenge are \emph{model-based} and \emph{parameter mapping} sonification (PMSon). Model-based techniques aim to transform datasets into a dynamic model, which one can interact with and aurally examine \cite{Hermann1999}. In contrast, PMSon exposes parameters that describe the data and maps these to sonic features.
Previous sonifications of molecular simulation data seemed to have favoured PMSon and \emph{auditory icons/earcons} rather than model-based techniques. This could be because molecular dynamics simulations already represent a physical model (see section \ref{ssec:compSim}) and adapting this dynamical system for the purposes of model-based sonification is challenging. 

  Rau et. al. \cite{Rau2016} demonstrate a PMSon for interrogating features of a static molecule in the \emph{Megamol} \cite{Grottel2015} visualization framework. Their approach was to create audio representations of features that are known to be chemically interesting, such as the forming and breaking of hydrogen bonds.
    \emph{Sumo} \cite{Grond2008} is a plug-in for the Python based molecular simulation environment \emph{PyMol} \cite{pymol}.This project had the fairly broad aim of providing a general framework for implementing various sonifications within \emph{PyMol}. For example parameterized earcons were utilized to represent pairwise distances and different conformations of amino acids. The designers hypothesized that users should be able to learn to distinguish between a set of 20 earcons (representing  different amino acids) and thus perceive conformational differences more readily than when the sonification track is absent.

  Hermann \cite{Hermann2001} describes some of the critical issues that arise when designing a PMSon, observing that mappings are not necessarily transparent to a first time listener without some kind of ``code book''. This point is reiterated by Wishart \cite{wishart2013} when describing the design of his piece, \emph{Supernova}, which sonified astronomical observations: ``...there is no particular reason to use one mapping rather than another. As a result, the sonic outcome would be entirely dependent on the mapping chosen."

  The issue of the arbitrariness of the chosen mappings is particularly interesting with regards to atomistic representations. Although they are physical, 3D objects, there is no way for us to experience them directly, either visually or aurally. Therefore there is an inevitable degree of flexibility in the design of any representation at this scale. This is illustrated by looking at the development of visualizations of molecular models which have showed a range of different approaches to representing molecular structure, from Linus Pauling's paper models of protein helices \cite{Pauling205}, to the metal, wood and plastic Kendrew models of protein structure \cite{kendrew1958}, through to the modern computer generated renderings in which a variety of different drawing schemes can represent the same structural information \cite{HUMP96}. 
  

  In terms of sonic design, there is no such established convention for representation. Depicting an atom as a sphere is, in some sense, an arbitrary decision, but there is an intelligible analogue in that they both have some sense in which they are spatially delimited. Attempting to define such a clearly delimited object in the audio realm is not as straightforward, neither spatially nor compositionally.  It is difficult to assert what constitutes a single atomistic object in a piece of sound design or music.

  There are a wide range of ``non-local" properties important in biomolecular science (e.g.\ potential energy, free energy, electrostatic energy, temperature, strain energy, conformational state membership, etc.). Such properties are extremely difficult to visualize using conventional rendering strategies (and even if there were effective strategies, would lead to significant visual congestion) owing to their non-locality. We believe that such properties are the most interesting to explore in a sonification context: hence our focus on free energy in this work.

  There is a question of the level of intervention that the sonification designer should take. If a dataset is rendered as directly as possible (i.e.\ converted to audio), then perhaps any audible features present must be features of the data. But this rule may depend on the source and type of data (as well as artifacts of the transform). For example, a set of measurements of how temperature changes over time might be treated differently to a non-local parameter that represents the overall instability of a system. In the latter case, it may be necessary to map the data in a less direct way to convey its provenance.  
  Scaletti \cite{Scaletti1994} categorises the directness of mappings through the idea of different orders: for $0$\textsuperscript{th} order, the data is directly read as an audio waveform, for $1$\textsuperscript{st} order the data is used to modulate an audio carrier signal.  
  Our sonification uses many 1st order, one-to-many \cite{Hunt2000} mappings that attempt to create a certain perceptual effect that relates to the significance of a given parameter.
  This approach may encounter a problem pointed out by \cite{Murphy2006}: that it is \emph{atheroetical} in that the decisions made are based on some subjective sound design process and the results often represent the designer’s sensibilities and preferences just as much as the underlying data set. The techniques used in this project are primarily parameter mappings, which certainly do encounter some of the issues raised above. Acknowledging these issues is important although addressing them all in detail is out of scope for this paper.

  This work extends the practice of molecular simulation sonification datasets by seeking an auditory display of the free energy landscape and its relation to fundamental dynamic processes of biomolecules, something which builds on our previous work developing real-time sonification strategies for molecular dynamics simulations (\cite{C4FD00008K, ds2016, glow2013}). To the best of our knowledge the methods outlined herein have not been attempted before. We model the underlying dynamics using two different types of Markov model, observed and hidden, from a set of simulation data of a simple biomolecule, Alanine dipeptide (AD). We extract features of the free energy landscape and the dynamical processes from the models and map these to sonic parameters. Simulated examples of the dynamics can then generate coupled visual and audio display of structural and dynamic information respectively. 

This paper is organized as follows: section \ref{sec:biodyn} explains some of the underlying physical ideas and the modeling of biomolecular dynamics, section \ref{sec:sonification} explains our sonification strategy, some details of the implementation are given in section \ref{sec:impl}, and our conclusions and outlook for further work are given in section \ref{sec:conclusions}. An example of the sonification described in this paper can be found at \url{https://vimeo.com/255391814}.

\section{Biomolecular conformational dynamics}
\label{sec:biodyn}
\subsection{Free energy landscape}

Biomolecules such as proteins and nucleic acids are dynamic objects comprised of $n$ atoms, each of which interacts with other atoms in the same molecule and to the cellular environment. Any given molecular system has $3n$ degrees of freedom, since any given atom can move in the $x$, $y$, and $z$ direction. Typically biomolecules are comprised of thousands of atoms, leading to high-dimensional dynamics.

At any given time, a molecule adopts a particular \emph{conformational} state. Researchers are typically interested in understanding the networks of conformational substates that characterize a particular molecule. Networks of highly connected states in which the system has a relatively long residence time are called \emph{metastable states}. Of particular interest in many applications is understanding how long it takes a molecule to travel between different metastable states. Conformational states are of interest because they are directly linked to the molecule's function. This picture \cite{Frauenfelder1598}  has been verified extensively through experiments \cite{Santoso715,Gebhardt2013} , theoretical and computational studies \cite{wales_2004, Buchete2008}.   

Any given conformational state is defined by its \emph{free energy}. Highly probable conformations have a lower free energy than improbable conformations. This rise and fall of free energy defines a \emph{free energy landscape} over the atomic coordinates which is illustrated in further detail in what follows.  

\subsection{Molecular dynamics simulation} \label{ssec:compSim}
The output of molecular dynamics (MD) simulations are a series of regularly timed snapshots (\emph{frames}) of atomic configurations as the system evolves, called \emph{trajectories}. An animated example trajectory can be viewed at \url{https://vimeo.com/255526473}.  With enough trajectories it is possible to understand the probability that a molecule occupies certain states and to develop a  map of the free energy landscape. However, calculating the free energy as a function of the $3n$ atomic coordinates would not provide meaningful insight into the system, owing to its high dimensionality. Researchers have therefore focused on ways to reduce the dimensionality of the system by identifying only those coordinates (or combinations of coordinates) which take the system from one metastable state to another.  Understanding which coordinates (among a large number of possibilities) resolve metastable states is an outstanding question in the study of chemical dynamics. 

\subsection{Markov Models}
Markov state models have found widespread use in recent years as a dimensionality reduction technique for analyzing the metastable dynamics of biomolecules \cite{CHODERA2014135}.  Their popularity stems from their ability to produce predictive and easy to understand results as well as their ability to parallelize the problem of resolving very long timescale processes.  There are two related Markov models in widespread use, observed Markov state models \cite{prinz2011markov} and hidden Markov models \cite{noe2013projected}. This work makes use of both of these models. 

Markov models transform a trajectory into a \emph{chain} of $n$ discrete states.  These states are called \emph{observed states} (or sometimes \emph{microstates}) and form the data from which both types of Markov model can be estimated. In general we refer to a chain as $x_{t}$ and a specific element by its position in the chain: $x_2 = 3$ denotes that the second frame of the chain is in state $3$.  We refer to the set of all possible discrete states as $\mathbf{x}$. 

An \emph{observed Markov state model} (or simply Markov state model, MSM) assumes  the probability of transitioning to observed state $b$ in a time $\tau$ given we are in state $a$,  $P(x_{t+\tau}=b|x_{t}=a)$, only depends on the states $a$ and $b$ and not on the states visited at times  $t-1, t-2,...,0$. This property is known as the \emph{Markov} property and any chain that satisfies this is known as Markovian. The dynamical information of the MSM is contained within a \emph{transition matrix}, $\mathbf{T}(\tau)$, whose elements are the conditional transition probabilities, i.e.\ $T(\tau)_{a,b} = P(x_{t+\tau}=b|x_{t}=a)$.  

The primary problem with observed Markov state models is that they contain too much information and it is typical to cluster the observed states into a smaller number of metastable states in order to make quantitative predictions about their dynamics. A \emph{hidden Markov model} (HMM) represents a sort of \emph{fuzzy} clustering of the observed states into  a set of metastable states (or hidden states, $\mathbf{X}$), i.e.\ instead of describing a particular conformation (observed state) as unambiguously belonging to a metastable state, a probability of membership is given. A HMM consists of a transition matrix for the  metastable states and a membership matrix $\mathbf{M}$ whose elements are the conditional probability of being in a metastable state ($A$) given a particular observed state ($a$), i.e.\ $M_{A,a} = P(X_{t} = A | x_{t} = a )$.

HMMs work well in describing biomolecular dynamics in the regime where the underlying dynamics are metastable.  In other words the proportion of observed states with membership probabilities intermediate between $0$ and $1$ are small in comparison with the total number of observed states. 

\subsection{Alanine Dipeptide Model}
\begin{figure}
\includegraphics[width=0.4\textwidth]{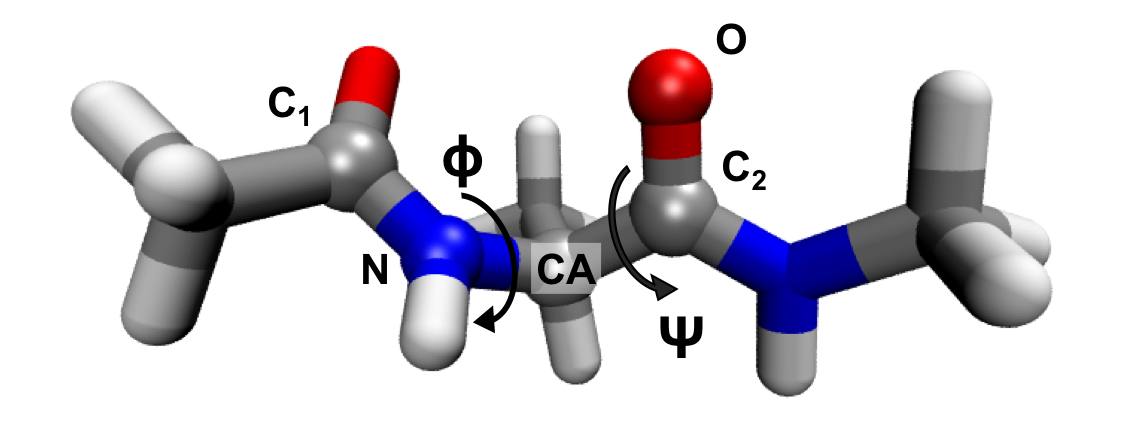}
\caption{Atomic structure of Alanine dipeptide (AD). The cylinders represent chemical bonds and their intersections represent atoms. Grey, blue, red and white colors are carbon, nitrogen, oxygen and hydrogen atoms respectively. The atoms involved in the $\phi, \psi$ dihedral angles are labeled and highlighted as spheres.  The $\phi$ angle is formed from the intersection of the planes formed by the atoms (C\textsubscript{1}, N, CA) and (N, CA, C\textsubscript{2}).  The $\psi$ angle is formed from the planes formed by the atoms (N, CA, C\textsubscript{2}) and (CA, C\textsubscript{2}, O).}\label{fig:ala2}
\end{figure}

\begin{figure*}
\includegraphics[width=\textwidth]{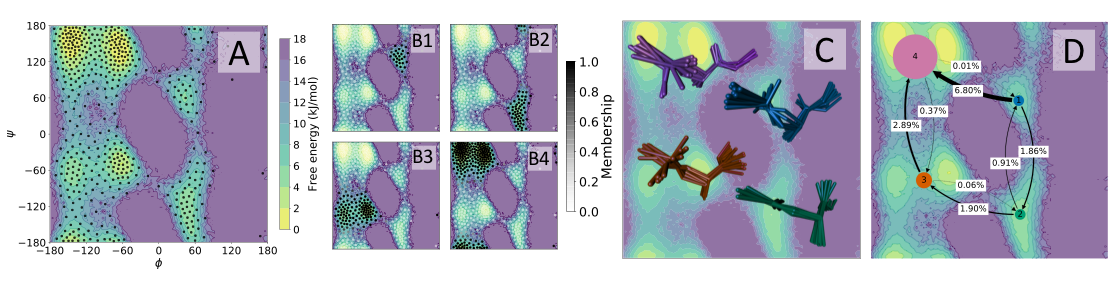}
\caption{The hidden Markov model for AD. \textbf{A}: The free energy landscape of AD projected onto the peptide $\phi$ and $\psi$ dihedral angles.  The lighter the colour the lower the free energy and hence the more stable those atomic configurations. The yellow regions define four metastable basins centered on coordinates $(60\degree,60\degree)$, $(60\degree,-120\degree)$, $(-120\degree,-60\degree)$, $(-120\degree,120\degree)$. The purple regions are not visited during the simulations used as input for the model. The black circles are the centers of 500 observed states for both the observed and hidden Markov models. \textbf{B1-4}: Each panel represents rows 1 - 4 of the membership matrix. Each circle represents an observed state coloured according to its membership probability to each metastable state. \textbf{C}: Ten sampled conformations of each metastable state of AD overlaid over the center of their free energy well. The hydrogen atoms have been removed for clarity. \textbf{D}: The transition matrix and stationary distribution of metastable states.  Each circle represents a metastable state with the area of each circle proportional to the stationary distribution. The arrows between each state show possible transitions with the width proportional to the conditional transition probability (also shown in the white boxes on each arrow). }\label{fig:hmm}
\end{figure*}

The ``hello world'' example of a biomolecule exhibiting metastable dynamics is Alanine dipeptide (AD), as shown in figure \ref{fig:ala2}. The metastable dynamics of AD are reasonably well described with reference to two dihedral angles made by atoms in the peptide bonds \cite{Chodera2007}, the $\phi$ and $\psi$ angles, also shown figure \ref{fig:ala2}.  The free energy landscape of AD projected onto these two dimensions is shown in figure \ref{fig:hmm}A. The light yellow colour denotes free energy \emph{wells}, i.e.\ regions with a low value of free energy which define the metastable states.  The lighter purple regions are those which are visited only briefly on the way to a metastable well, known as transition regions. The free energy landscape was discretized into $500$ observed states, each of which has a centroid shown by the black circles. Each frame of the trajectory is assigned to the nearest observed state. As the dihedral angles are periodic, conformations with $\phi/\psi=180\degree$ are equal to those with $\phi/\psi=-180\degree$.  This means that rather than a 2D plane, the free energy landscape actually resides on a \emph{torus}, i.e.\ each edge of the chart should be wrapped around to meet the opposite side. For the sake of simplicity we show it here in the form in which it is typically rendered by practitioners in the field. 

For the purposes of this paper two models were created - an observed MSM and a HMM. Details of the data and calculations used to generate the models can be found in section \ref{sec:impl}. The estimated transition matrix for the MSM results in $500$ eigenvectors which describe the various relaxation modes of the dynamics.  The first eigenvector $\mathbf{q}^{1}$ is equal to the stationary distribution, $\mu(\mathbf{x})$.  The next three eigenvectors are slow relaxation modes ($\mathbf{q}^{2,3,4}$) which define population transfer \textit{between} metastable states.  The next five eigenvectors ($\mathbf{q}^{5-9}$) are fast relaxation modes which define population transfer \textit{within} metastable states. Each relaxation mode has an associated timescale (the corresponding eigenvalue). The remaining eigenvectors were discarded as the associated timescales  for these modes was faster than the time resolution of the data used to estimate the model and so were not considered statistically robust. The HMM was estimated by assuming four metastable states. 

The results are shown in figure \ref{fig:hmm}.  Figure \ref{fig:hmm}D shows the metastable state transition matrix, $\mathbf{T}$. Each circle represents a metastable state with the area of the circle proportional to its stationary distribution, $\mu(X_{1,...,4})$.  The arrows show the conditional probability of transitioning to each state. For example, the probability of transitioning from state $1$ to state $4$ is $6.79\%$.  State $4$ is by far the most stable, followed by $3$, $2$ and then $1$. As there are no transition regions between state $1$ and $3$ and between $2$ and $4$, the probability of transitions between these pairs of states is zero. Figure \ref{fig:hmm}C shows an overlay of ten characteristic structures for each metastable state, overlaid over their respective free energy wells. Figures \ref{fig:hmm}B1-4 show the rows of the membership matrix. Each circle represents one of the observed states, coloured according to the membership probability to each metastable state.  The  partitioning of the basins is clearly shown by the regions of black circles (high probability of metastable membership) vs. white circles (low probability of metastable membership). 

\section{Sonification} \label{sec:sonification}

\begin{figure}[th]
\includegraphics[width=0.4\textwidth]{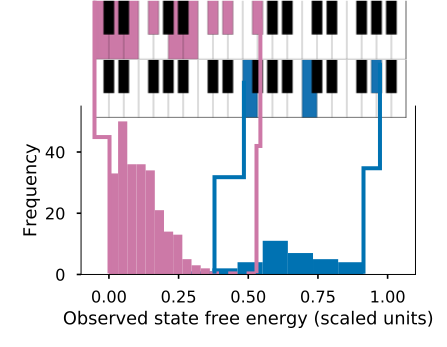}
\caption{The mapping of static properties of the metastable states to note clusters. Each of the observed states was first assigned to exactly one metastable state. The bottom chart shows the distribution of free energies of observed states which have been assigned to the metastable states $1$ (blue) and $4$ (pink). States $2$ and $3$ are not shown for clarity.  The coloured notes of the keyboard are the note clusters used to represent the metastable states. The relative upper and lower bounds of the distributions determine the highest and lowest notes in the cluster (as shown by the vertical connectors).  The relative area of each distribution determines the number of notes in the note cluster. The ratio of the areas of state $4$'s distribution to state $1$'s distribution is approximately $3:1$. This determines the $9:3$ ratio of the number of notes in each note cluster.}\label{fig:staticprop}
\end{figure}

\begin{figure}[t]
\includegraphics[width=0.4\textwidth]{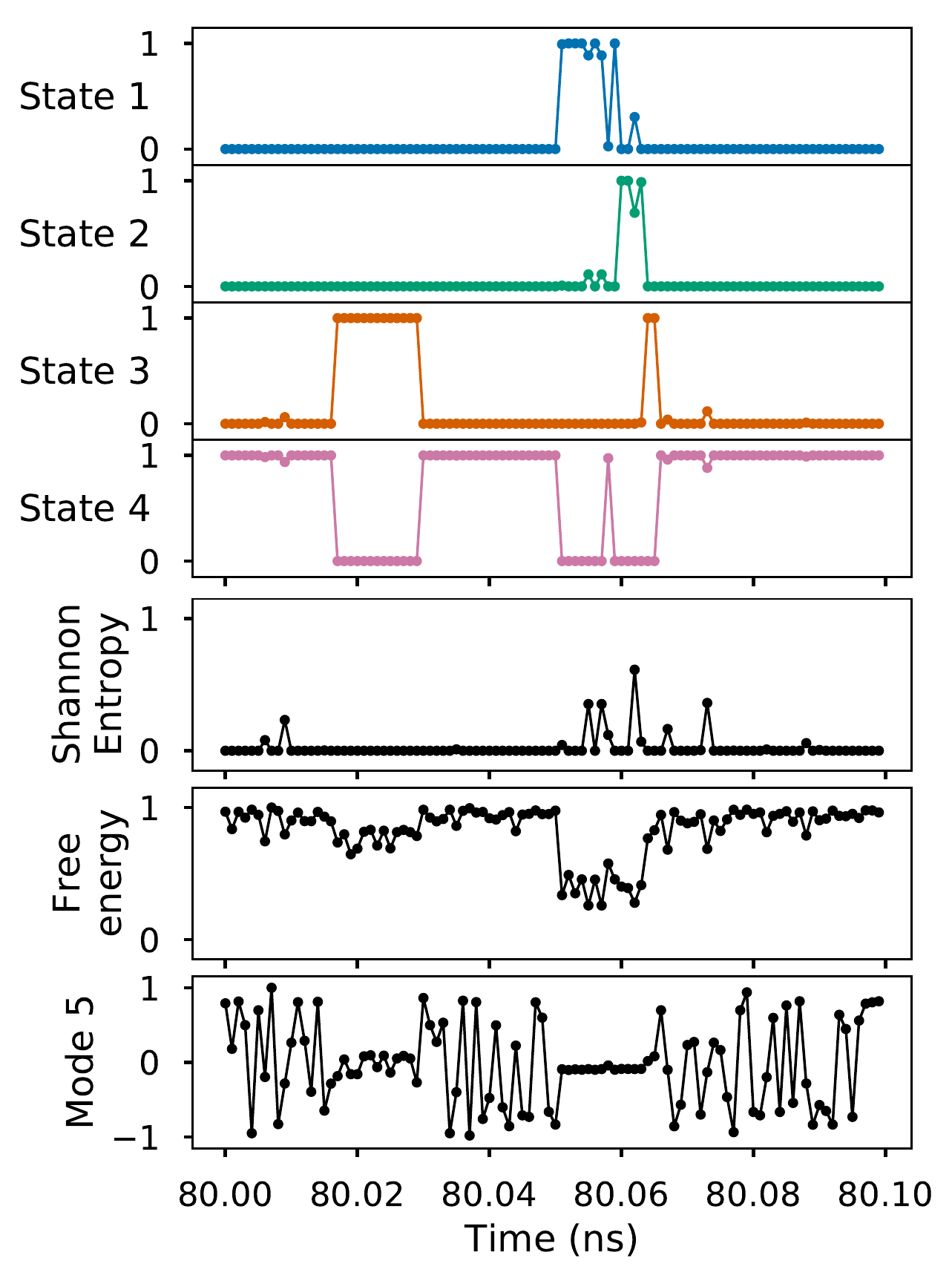}
\caption{The dynamic parameters of the model. Each panel shows how a model parameter varies over $0.1$ns (100 frames) of the example trajectory. \textbf{State 1-4} show the membership probability of each frame's observed state to each metastable state. The values in each frame across all four panels sum to 1. \textbf{Shannon entropy} measures the uncertainty of the assignment of each observed state to the metastable states.  High entropy indicates an observed state could be considered to belong to more than one metastable state. \textbf{Free energy} is the free energy of each observed state, the lower the free energy the more stable the state.  \textbf{Mode 5} is the first fast relaxation mode which redistributes population within a metastable state. The other four fast modes are not shown.}\label{fig:dynprop}
\end{figure}

The purpose of the sonification is to provide information on features of the free energy landscape simultaneously with visual display of structural information, i.e.\ a molecular animation from an example trajectory.  As this trajectory traverses the different parts of the free energy landscape the sonification brings out three of its features with the following distinct layers in the audio design: (1) a continuous pad sound that represents membership of different metastable states and their properties, (2) a pulse sound that represents the stability of the system and, (3) a set of synthesized tones that represent how the system is changing within each metastable state.

The sonification therefore requires three objects: a model of AD dynamics, an example trajectory and an animation of the example trajectory. In this work the data used to estimate the model of AD (the input trajectories) are different to the example trajectory although in principle trajectories from the input data could be used as example trajectories. 

\subsection{Sonification parameters}
There are two categories of sonification parameters: \emph{static} and \emph{dynamic}. Static parameters do not change over time and are initialized at time $0$. They are derived from the model parameters alone. Dynamic parameters are those derived from each frame (i.e.\ from each observed state) of the example trajectory as well as the model parameters.
\subsubsection{Static parameters}\label{ssec:static}
The static parameters are all derived from the shape of the free energy wells associated with each metastable state. Each part of the free energy landscape is assigned a probability of membership to a metastable state so the limits of each well are not well defined. In order to overcome this problem each observed state is assigned to the metastable state for which it has the highest probability of membership.  

The free energy for each observed state $a$, $F(a) = -kT\ln(\mu([\mathbf{x}]_{a}))$ was calculated and scaled to lie in the range $(0,1)$. Here $k$ refers to the \emph{Boltzmann} constant and $T$ to the temperature. The static parameters for each metastable state were derived from a histogram of the free energies of observed states assigned to that metastable state.  We denote the histogram for metastable state $A$ as $h_{A}(F)$ (or $h$ in general). The properties we derive from $h(F)$ are (1) its upper bound, $UB[h]$, (2) its lower bound $LB[h]$ (3) its area $A[h] = \int^{1}_{0}\mathrm{d}F h(F)$. 

The histograms for states $1$ (blue) and $4$ (pink) are shown in figure \ref{fig:staticprop}. The model has four metastable states meaning there are $4\times 3 = 12$ static parameters. The upper and lower bounds are related to the free energy well minima and maxima for each metastable state.  The area of the histograms is proportional to the overall volume of the free energy well.

\subsubsection{Dynamic parameters}
The dynamic parameters change with each observed state in the example trajectory. For each observed state these parameters are: (1) its \emph{probability of membership} to each of the four metastable states, (2)  the \emph{Shannon entropy} of its metastable state assignments (3) its absolute \emph{free energy}, (4) its \emph{projection} into the five fast relaxation modes. For each observed state there are $4+1+1+5 = 11$ dynamic parameters.  

The membership probability describes the probability that a given observed state can be assigned to a given metastable state.  For observed state $a$ there are four membership probabilities given by $([\mathbf{M}]_{1,a}, [\mathbf{M}]_{2,a},[\mathbf{M}]_{3,a},[\mathbf{M}]_{4,a} )$.  

The information or Shannon entropy is a measure of the degree of certainty with which the assignment of an observed state to a particular metastable state can be made.  The Shannon entropy for an observed state $a$, $H_{a}$ is given by $
H_{a} = -\sum^{4}_{i=1} [\mathbf{M}]_{i,a} \ln([\mathbf{M}]_{i,a})$. 
$H_{a} = 0$ indicates the observed state is definitely in one metastable state.  $H_{a} = \ln(4) \approx 1.4$ indicates it is equally likely to be in any of the four metastable states.  

The free energy of observed state $a$, $F(a)$ measures the observed state's global stability. These were the same free energies used in the calculation of the static parameters in section \ref{ssec:static}. 

The projection of observed state $a$ onto the $i$'th fast relaxation modes is given by $[\mathbf{q}^i]_{a}$. Large oscillations in a projection indicate that the system is relaxing along that mode.  The values of these projections were scaled to lie in the range $(-1,1)$. Figure \ref{fig:dynprop} shows the dynamic parameters for a section of an example trajectory. 

\subsection{Mapping} \label{ssec:map}

As well as categorizing the parameters as static or dynamic, a further distinction is drawn between those that relate to changes within the current metastable state (intrastate) and those that pertain to changes between the metastable states (interstate). Table \ref{tab:table1} summarizes these two categorizations.  This second classification is useful because it draws the distinction between parameters based on features of the physical dynamics rather than how they were generated. The interstate parameters represent the most physically important features of the dynamics and free energy landscape and so form the core of the mapping strategy.  

\subsubsection{Interstate parameters}
The three static parameters for each metastable state, $UB[h]$, $LB[h]$, $A[h]$ are interstate parameters. These features of the model are principally related to two physical features: the relative stability and conformational flexibility of each metastable state. The focus of the mapping strategy is to find a way to describe these features aurally. Note clusters are chosen to be associated with each metastable state. The reason underlying this choice arises from the fact that transitioning between tonal groupings is a common musical device; it is hypothesized that non-trained listeners should be able to perceive the changes in note density and relative pitch. 

The maximum note range is predefined to three octaves. The relative values of $A[h]$ for each metastable state defines the number of notes in its corresponding note cluster, these notes are then evenly distributed between the lowest and highest notes. The values of $LB[h]$ and $UB[h]$ define the lowest and highest notes of each cluster respectively. This is shown in figure \ref{fig:staticprop} for metastable states $1$ and $4$. State $4$ (pink) has a smaller range, but a larger area resulting in a dense, tightly spaced note cluster at the lower extent of the note range. State $1$ (blue) has a large range but small area resulting in a more sparse note cluster.  State $4$ has its lower bound below that of state $1$ and so the lowest note of the cluster is below that of state $1$. An accepted limitation of this mapping is that it does not attempt to classify and choose the note clusters in terms of their harmonic relationships. Instead, it deals with them as distributions of notes within a range, with a given extent and density. It may be possible to use the work of Lewin \cite{Lewin2007} and others to create a hierarchy of tonal groupings from which to choose but this is highly genre specific. This is an outstanding issue in that listeners may interpret the harmonic relationship of two note clusters as significant when this is not intended as part of the mapping (e.g.\ stacked whole tones vs. stacked fourths).     

The membership probabilities ($\mathbf{M}$) control the choice of note cluster by linearly interpolating between the values defined by the static parameters of each state. This means that if $[\mathbf{M}]_{1,a}= 1$ then the cluster defined by the static properties of metastable state $1$ will be used (the same follows for all the states). In the case that $[\mathbf{M}]_{1,a}= [\mathbf{M}]_{2,a}=0.5$ then the lowest and highest notes will lie halfway between those defined by the static parameters of state $1$ and $2$. 

Large values of the Shannon entropy ($H$) represent observed states which could be assigned to more than one metastable state, physically this means they are in transition regions between two metastable states. $H$ is mapped to the width and rate of frequency modulations of the voices for the note clusters as well as the bandwidth of the filter such that it tends toward noise for higher values of $H$. This is a one-to-many mapping that is designed to create a perceptual effect of instability in the pad sound. $H$ tends to remain at $0$ with occasional spikes in amplitude as the system enters a transition region. This is shown in figure \ref{fig:dynprop} where there are spikes between $80.05$ and $80.06$ nanoseconds as transitions occur between all four states. These spikes, although fleeting, are important features. In order to render them as noticeable a smoothing process for decreasing values is employed such that incoming values decrease slowly but can increase quickly. Figure \ref{fig:synthProcess} shows the structure and mapping for a single synthesiser voice used in the creation of the pad sound (there are 10 voices used in total). The mappings are designed to distinctly represent each metastable state and to indicate when a transition between states occur.

\subsubsection{Intrastate parameters}
The absolute free energy ($F(x_t)$) of each observed state measures its stability relative to the global free energy minimum. This parameter allows the sonification to draw a distinction between an observed state being globally unstable and yet part of a relatively stable metastable state (or vice versa). 

$F(x_t)$ is used to modulate a pulse sound that underpins the whole sound world. The rationale here is that a pulse or kick drum type sound is commonly used as a fundamental of a sound world upon which the other elements are constructed. This mirrors the way the free energy landscape underpins the dynamics of the physical system. The variable is reversed in terms of its polarity and transformed such that decreases in its value create an increase in tempo and a brightening of the tone of the pulse. 

The fast mode projections are oscillatory signals between -1 and 1 and are exported as PCM wav files (a $0^{\mathrm{th}}$ order mapping in Scaletti's terminology \cite{Scaletti1994}). 
At the 20Hz frame-rate being used the oscillations are subsonic. \emph{Scanned synthesis} is used in order to render the content as audible \cite{Verplank2001}. This method was developed to allow for the direct manipulation of synthesis timbre using a physical model and is essentially an extension of wavetable synthesis. In this case it allows for a rolling window of the audio buffer to be scanned at a given frequency, with the effect that increased frequency and amplitude of oscillation in the window results in a brighter timbre. These can be heard panned from hard left (mode $\mathbf{q}^{5}$) to hard right (mode $\mathbf{q}^{9}$) and the scanning frequency is defined by the first five notes of the currently defined note cluster.
\renewcommand{\arraystretch}{1.2}
\begin{table}
  \begin{center}
    \begin{tabular}{l|l|l|l}
      \textbf{Type} & \textbf{Scope} & \textbf{Parameter} & \textbf{Mapping}\\
      \hline 
      Dynamic & Intra & Free Energy \big($F(a)$\big)             & Pulse sound\\         
      		  &       & Fast Mode \big($[\mathbf{q}^i]_{a}$\big) & Scanned synth tones \\
              \cline{2-4}
              & Inter & Shannon Entropy \big($H_{a}$\big)        & Pad sound  \\
              &       & Membership                               & Chosen cluster \\
              &       & probability \big($[\mathbf{M}]_{a}$\big) & \\
      \hline
      Static  & Inter & Well min \big($UL[h]$\big)               & Lower note  \\
      		  &       & Well max \big($UB[h]$\big)               & Upper note  \\
      	      &       & Histogram area \big($A[h]$\big)          & Number of notes \\
    \end{tabular}
    \caption{Simulation parameters exposed via OSC. \textbf{Type} refers to whether the parameters are fixed at time $0$ (static) or change with each trajectory frame (dynamic). \textbf{Scope} refers to whether the parameters are describing transitions between states (inter) or within a state (intra).}
    \label{tab:table1}
  \end{center}
\end{table}
\begin{figure}
\includegraphics[width=0.5\textwidth]{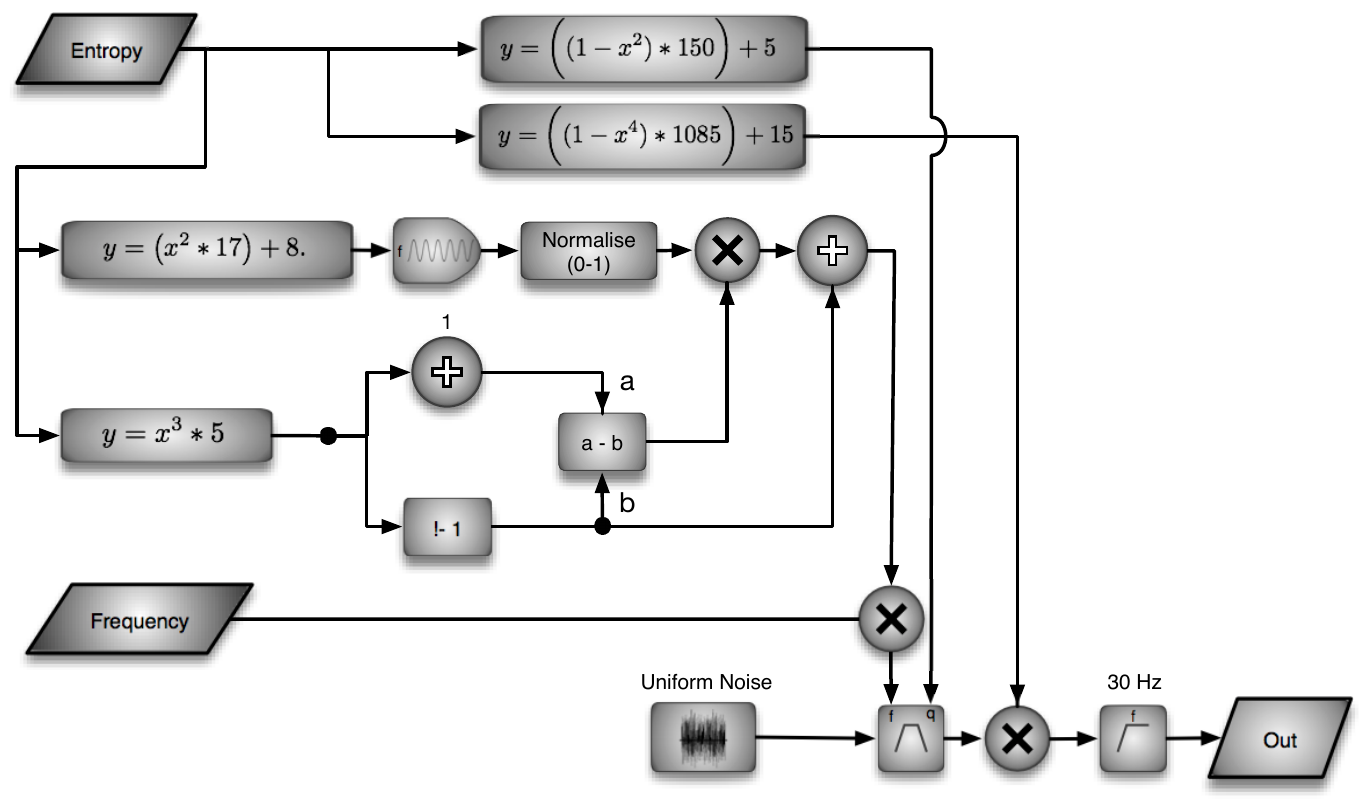}
\caption{Single synthesis voice as used for each note of a metastable state note cluster. The set of multiple instances (one per note) form the pad sound. The inputs are the frequency of each note in the note cluster and the entropy. The note cluster is determined by the membership probability of the current frame. The entropy is shaped, scaled and offset in various ways in preparation for use as a modulation source.}\label{fig:synthProcess}
\label{fig:entropymap}
\end{figure}

\section{Implementation} \label{sec:impl}
The simulation data was taken from a publicly available repository which accompanies the paper \cite{Chodera2007} found at \url{https://simtk.org/projects/alanine-dipeptide/}.  A full explanation of the methods used to generate the data are given in the paper.  All modeling was done using Python 3.5 programming language, the Markov models were generated by PyEMMA 2.4 \cite{scherer_pyemma_2015}. Both the HMM and MSM used the same set of 500 observed states and each were estimated with a lag time of $\tau=1.0$ps. The HMM was estimated by specifying four metastable states. This number was chosen so that the approximation of metastability was most accurate at the lag time used. The molecular animations were generated using VMD \cite{HUMP96}.  

A $500$ns long example trajectory was used in the sonification.  This was generated from the HMM, rather than using an input trajectory as they were all of insufficient length to sample each metastable state regularly.  While this trajectory does not strictly obey the original equations of motion used to generate the input data it reproduces all the modeled features (transition probabilities, relaxation timescales, stationary distribution) and so is indistinguishable from a trajectory generated using the original equations of motion. 

A Python script, using the package OSC 1.6.4, was used to create a client which sent the static and dynamic parameters as messages to the audio processing software. The messages were sent using the OSC protocol. The dynamic parameters were sent at a rate of 20 trajectory frames per second (corresponding to a ratio of $20$ps to $1$s simulation to physical time). The audio processing was implemented in Max/MSP.
 
Everything required to reproduce this work (except the input data which can be downloaded separately) can be found at \url{https://osf.io/rzp3k/?view_only=b5802dfce6da4dd59dfb6b406ae033f0}.  

\section{Conclusions and Future work}\label{sec:conclusions}

We have presented a strategy for mapping important features of the dynamics and the free energy landscape of Alanine dipetide (AD) to sonic parameters to create an auditory display.  This auditory display can be used in tandem with visual display techniques to help build an intuitive understanding of how the physical structure of AD relates to the underlying free energy surface and the resulting dynamic processes. Our method and implementation  now requires user testing to ascertain how well sonic representations of this sort convey the desired information when used by chemists in a research or teaching context. A further question  is whether sonifying additional model parameters adds to our understanding of biomolecular dynamics. 

In the future we hope to extend this implementation to allow a degree of interactivity in manipulating the example trajectory. Initially, this would take the form of allowing the user to manipulate the playback position, speed and loop points of the trajectory. This would allow them to focus on regions of interest in the dynamics. In the long term, allowing the user to manipulate the example trajectory using interactive molecular dynamics  (for example, NanoSimbox \cite{2018arXiv180102884C}) and hear the resultant sonic effects is an exciting prospect and opens up the possibility of using the system as an instrument for musical expression as well as a data exploration tool.   

\section{ACKNOWLEDGMENT}
The authors would like to thank Lars Bratholm for his insightful input in developing a probabilistic approach to the mapping strategy. 
David Glowacki acknowledges funding from the Royal Society (UF120381), the EPSRC (impact acceleration award and institutional sponsorship award), and the Leverhulme Trust (Philip Leverhulme Prize). Robert Arbon is funded by a studentship from the Royal Society (RG130510), with additional support from the Leverhulme Trust. Alex Jones is jointly supported by a studentship from the EPSRC and Interactive Scientific Ltd. Lars Bratholm acknowledges funding from EPSRC Programme Grant EP/P021123/1.
\label{sec:ack}

\bibliographystyle{IEEEtran}
\bibliography{refs2018}
%
%
%
%

\end{sloppy}
\end{document}